\documentclass[fleqn,usenatbib]{mnras}

\usepackage{newtxtext,newtxmath}
\usepackage{mathtools}
\usepackage[T1]{fontenc}
\DeclareRobustCommand{\VAN}[3]{#2}
\let\VANthebibliography\thebibliography
\def\thebibliography{\DeclareRobustCommand{\VAN}[3]{##3}\VANthebibliography}

\usepackage{tikz,xcolor,hyperref}

\definecolor{lime}{HTML}{A6CE39}
\DeclareRobustCommand{\orcidicon}{
	\begin{tikzpicture}
	\draw[lime, fill=lime] (0,0) 
	circle [radius=0.16] 
	node[white] {{\fontfamily{qag}\selectfont \tiny ID}};
	\draw[white, fill=white] (-0.0625,0.095) 
	circle [radius=0.007];
	\end{tikzpicture}
	\hspace{-2mm}
}

\foreach \x in {A, ..., Z}{\expandafter\xdef\csname orcid\x\endcsname{\noexpand\href{https://orcid.org/\csname orcidauthor\x\endcsname}
			{\noexpand\orcidicon}}
}

\usepackage{graphicx}	
\usepackage{amsmath}	

\usepackage{hyperref}
\hypersetup{
  colorlinks   = true, 
  urlcolor     = blue, 
  linkcolor    = blue, 
  citecolor   = blue 
}
\usepackage{xcolor}
\usepackage{multirow}
\usepackage{subcaption}
\usepackage{ulem}
\usepackage{array,multirow}

\newcommand{\h}{$^{\rm h}$}
\newcommand{\m}{$^{\rm m}$}
\newcommand{\s}{$^{\rm s}$}
\newcommand{\hb}{H$\beta$}
\newcommand{\oiii}{[O\,{\sc iii}]}

\newcommand{\hnii}{{\rm H}$\alpha$+[N~{\sc ii}]}
\newcommand{\ha}{H$\alpha$}
\newcommand{\sii}{[S~{\sc ii}]}
\newcommand{\nii}{[N~{\sc ii}]}
\newcommand{\sulfur}{[S~{\sc ii}]}
\newcommand{\nitrogen}{[N~{\sc ii}]}
\newcommand{\oxygen}{[O~{\sc iii}]}
\newcommand{\flux}{$10^{-15}$ erg s$^{-1}$ cm$^{-2}$}

\title[Optical emission from SNR G7.7$-$3.7]{Detection of optical emission from the supernova remnant G7.7$-$3.7\footnote{Based on observations made with the Isaac Newton Telescope operated on the island of La Palma by the Isaac Newton Group of Telescopes in the Spanish Observatorio del Roque de los Muchachos of the Instituto de Astrofísica de Canarias.}}

\author[Dom{\v c}ek et al.]{V. Dom{\v c}ek$^{1,2\orcidA}$\thanks{E-mail: vdomcek@gmail.com},
J.~V. Hern\'andez Santisteban$^{3\orcidB}$,
A. Chiotellis$^{4,5\orcidC{}}$,
P. Boumis$^{4\orcidD{}}$,
\newauthor
J. Vink$^{1,2,6\orcidE}$,
S. Akras$^{4\orcidF{}}$,
D. Souropanis$^{4,7,8\orcidG{}}$,
P. Zhou$^{1,9\orcidH}$
and A.~de~Burgos$^{8,10,11\orcidI}$
\\
$^{1}$Anton Pannekoek Institute for Astronomy, University of Amsterdam, Science Park 904, 1098 XH Amsterdam, The Netherlands\\
$^{2}$GRAPPA, University of Amsterdam, Science Park 904, 1098 XH Amsterdam, The Netherlands\\
$^{3}$SUPA Physics and Astronomy, University of St Andrews, St. Andrews, KY16 9SS, Scotland, UK\\ 
$^{4}$Institute for Astronomy, Astrophysics, Space Applications and Remote Sensing, National Observatory of Athens, Penteli, Greece\\
$^{5}$$4^{\rm th}$ Lykeion Acharnon, Ag. Petrou and Echinou, 136 74 Acharnes, Greece\\
$^{6}$SRON, Netherlands Institute for Space Research, Utrecht, The Netherlands \\
$^{7}$Department of Physics, National and Kapodistrian University of Athens, Panepistimiopolis, 15784 Zografos, Greece\\
$^{8}$Isaac Newton Group of Telescopes, Apartado 321, E-38700 Santa Cruz de La Palma, Canary Islands, Spain\\
$^{9}$School of Astronomy and Space Science, Nanjing University, Nanjing 210023, People’s Republic of China\\
$^{10}$Universidad de La Laguna, Dpto. Astrof\'isica, E-38206 La Laguna, Tenerife, Spain \\
$^{11}$Instituto de Astrof\'isica de Canarias, Avenida V\'ia L\'actea, E-38205 La Laguna, Tenerife, Spain
}

\date{Accepted XXX. Received YYY; in original form ZZZ}

\pubyear{2023}

\begin{document}
\label{firstpage}
\pagerange{\pageref{firstpage}--\pageref{lastpage}}
\maketitle

\begin{abstract}
We present the first optical study of the supernova remnant (SNR) G7.7$-$3.7, 
with the aim of determining its evolutionary phase since it has been suggested 
to be the remnant of SN~386~AD.
We obtained narrow-band images in the filters \hnii, \hb, \oiii, \sii that revealed faint optical emission in the southern region of the SNR consisting of two filaments elongated in the east-west direction aligned with the X-ray emitting region of the remnant. The filaments were seen in \hnii, \oiii\ images and marginally in the \sii\ images, with a non-detection in \hb. Long-slit spectroscopy of three regions along one filament revealed large ratios of \sii/\ha \ =$ (1.6-2.5)$, consistent with that expected for a shock-heated SNR. 
The \sii\ doublet ratio observed in two of the regions implies an upper limit for the electron density of the gas, with estimates falling below {$400$~cm$^{-3}$} and {$600$~cm$^{-3}$} in the respective areas.
We discuss potential physical mechanisms that formed the observed optical filaments and we suggest that most likely they resulted by a collision of the SNR with a dense circumstellar shell lying at the southern region of the remnant.
\end{abstract}

\begin{keywords}
ISM: G7.7-3.7 -- ISM: supernova remnants -- supernovae: SN 386 -- Shock waves -- Techniques: photometric, spectroscopic
\end{keywords}



\section{Introduction}

Supernova remnants (SNRs) result from the interaction between an interstellar, or circumstellar, medium and a shock-wave, 
initiated by energetic supernova (SNe) that mark the death of massive stars or by the thermonuclear combustion of white dwarfs in interactive binaries.
They are regarded as factories of cosmic rays in our Galaxy, the hosts of young neutron stars, and are essential for studies of astrophysical shocks \citep[see][for a review]{Vink2020Book}. Optical emission, in particular, plays an essential role in the studies of SNRs as most of the extragalactic SNRs have been identified and characterised only using optical observations \citep{Leonidaki2013,Long2017}.
The optical band plays a less dominant but still an important role in the study of Galactic SNRs. It provides vital constraints on the ambient-medium densities and shock speeds of two distinct regimes of astrophysical shocks - radiative and non-radiative. 

Radiative shocks exist in regions of SNRs where the shock has significantly slowed down ($\lesssim200$~km s$^{-1}$). In the optical range, this leads to the formation of prominent H$\alpha$, \sii, \oiii\ and also \nii\ emission lines. Due to the necessity of low shock velocities, radiative shocks are mostly associated with evolved SNRs, such as N49 \citep{Bilikova2007} or HB3 \citep{Boumis2022}, but are not exclusive to them. Younger SNRs like Kepler \citep[see][for a review]{Vink2016}, Cas A \citep{Milisavljevic2013}, and even SN 1987A \citep{Mccray2016} also exhibit radiative optical emission originating in dense clumps of circumstellar material (often rich in nitrogen) or ejecta (rich in carbon-burning products or intermediate-mass elements), where the shock has locally slowed down due to the presence of higher density gas. 

Non-radiative shocks are formed when the cooling timescale of the shocked gas is substantially larger than the dynamical timescale of the forward shock expansion. These are found in SNRs that are characterised by relatively high velocities ($\gtrsim200$~km~s$^{-1}$) propagating in a moderately low density environment. Consequently, non-radiative shocks are found in young \citep[e.g. Tycho's SNR;][] {Ghavamian2000} or in evolved SNRs that have maintained their high speed due to their propagation in a low density ambient medium \citep[e.g. Cygnus Loop;][]{Hester1994}. Non-radiative shocks are often accompanied by pure Balmer-line emission from the SNR shock fronts, seen predominantly in the early phases of the SNR evolution and are almost exclusively present in young Type~Ia SNRs (e.g., SN1006, SNR 0509-67.5). The emission is observed in very thin filaments originating in regions with higher shock speeds \citep{Heng2010} which are thought to be important sites for cosmic-ray acceleration \citep{Chevalier1980,Vink2010,Vink2013,Morlino2013}.

Historical supernova remnants --- e.g., SN1006, Tycho and Kepler --- form only a small fraction of the SNR population, and have been well studied at optical wavelengths. Recently, \cite{Zhou2018} suggested that G7.7$-$3.7 (PKS 1814-24) belongs to this category, being the potential SNR of SN~386~AD. Moreover, G7.7$-$3.7 might be one of a few known historical SNR that resulted from a very low-luminosity SN and, therefore, could be of particular value for studying the evolution and origin of this small group of SNe \citep[$<$5$\%$ of all Type II SNe; ][]{Pastorello2004}. Recent studies found that SN~1181 was also a sub-luminous Type Iax historical supernova \citep{Schaefer2023}, and its probable remnant is a newly identified nebula called Pa~30 \citep{Ritter2021}.

G7.7$-$3.7 has not been extensively studied in the past, but the suggestion that G7.7$-$3.7 may be the remnant of a low- luminosity historical supernova, heightens the need to study this SNR more closely. G7.7$-$3.7 was first listed as a radio source with a linear polarisation of $\sim10\%$ in early catalogues from 1960s \citep{Gardner1969}. The high degree of polarisation was an early indication that G7.7$-$3.7 is an SNR. Subsequent targeted radio investigations \citep{Milne1974,Milne1986,Dubner1996} revealed an extended object with a shell-like structure, with angular diameter of $\sim 22^\prime$ and the overall radio spectral index $\alpha \sim 0.32$, flatter than the average index for SNRs \citep[see][figure~12.2]{Vink2020Book}. The distance has been suggested to be $4.5\pm1.5$~kpc  based on the $\Sigma-$D (surface brightness - diameter) relation \citep{Milne1986,Pavlovic2014,Green2015}. The visual extinction as measured from X-ray data was found to be relatively low, A$_\mathrm{v}\sim 1.2 \pm 0.2$ \citep{Zhou2018}.

So far, investigations of G7.7$-$3.7 in the optical \citep{VandenBergh1978} or $\gamma$-rays \citep{Acero2016} did not yield the detection of a SNR. In mid-infrared, three dust regions with temperatures of 50-100~K were identified \citep{Milne1986,Arendt1989}. The brightest one is centred on a planetary nebula in the field of view of the SNR, the second is coincident with the location of radio bright and steep feature ($\alpha \sim -0.6$) in the west of the SNR, and the third lies just outside of the radio shell in the southeast of G7.7$-$3.7  \citep{Milne1986}.
Later studies in X-rays \citep{Giacani2010,Zhou2018} have shown an X-ray counterpart to the radio structure in the south of the SNR, and a point-like source in the northeast \citep{Giacani2010}. An X-ray analysis of the southern shock region indicates the age of the plasma of be $t\sim1.2\pm0.6$~kyr,
based on the ionisation age 
\citep{Zhou2018}, with largely sub-solar abundances indicating a shocked interstellar medium. 

In this work, we report on deep targeted optical observations of SNR G7.7$-$3.7. {In particular, we obtained images using narrow-band filters covering the emission lines of \hnii, \hb, \oiii, \sii\ and low resolution spectra covering the range of 3500 \AA\ -- 9800 \AA.} In section \ref{g7_sec:data_analysis}, we describe the data reduction process, while in section \ref{g7_sec:results_discussion} we report on our findings and discuss potential implications for the SN~386~AD origin scenario.

\section{Observations and analysis}
\label{g7_sec:data_analysis}

\subsection{Imaging} 

We observed SNR G7.7$-$3.7 on 22-25 August 2019 (program ID:ING.NL.19B.006) 
and on 23 June 2022. The observations were performed
using the Wide Field Camera (WFC) instrument, at the prime focus of the 2.5m Isaac Newton Telescope (INT), in La Palma, Spain. The field of view of the detector is $\sim 34^\prime \times 34^\prime$ with a pixel scale of $0.333^{\prime\prime}$ pix$^{-1}$. In 2019 we obtained data in the narrow-band filters \hnii, \hb, \oiii, \sii\ and the SDSS broad-band filters $r^\prime$, $g^\prime$, which were used for continuum subtraction. In 2022 we expanded the data-set with additional observations with the \hb\ and $g^\prime$ filters. The observations were performed in a dither pattern to fill the gaps between its four chips. Sky flat fields were taken nightly during the morning and evening twilight.  

The 2019 data were pre-selected based on the average FWHM and the number of stars detected in each image (see Table~\ref{g7_tab:observation_log} for details on image quality per filter). Several additional images had to be excluded due to unexplained anomalous behaviour of the detector. This resulted in combined exposure times in the narrow-band filters \hnii, \hb, \oiii, \sii\ of 90, 15, 37, 30 minutes respectively, and two broad-band filters, $r^\prime$ and $g^\prime$, with exposures times of 5 minutes. 

Data reduction of the 2019 data was performed with the {\sc THELI~(v2)} software \citep{Erben2005,Schirmer2013}, using the default settings with exceptions in background modelling\footnote{Recommended setting of the \texttt{ focal plane array handling} parameters were for Gaia DR1 and WFC detector in {\sc THELI (v2)} following: \\ Stability type - instrument; Mosaic type - same crval; FPA mode - new}. We used the option of \texttt{smoothing scale for background model} with a set value of 300px and changed the co-addition setting to the median. We used the Gaia DR1 catalogue \citep{Gaia2016a,Gaia2016b} within {\sc THELI} with a magnitude limit of 18~mag in order to improve our astrometric solution of the field. We did not use the {\sc THELI} sky-background subtraction option at this stage. For further analysis, we reprojected all images to a common world coordinate system using the \hnii\ image as a reference. We employ \texttt{reproject\_interp} of the {\sc astropy} affiliated library {\sc reproject}, which aligns and reprojects data to a new common pixel grid using interpolation in order to match physical pixel coordinates between the images \citep{AstropyCollaboration:2013}.
Based on the examination of the {\sc THELI}-produced count-rate mosaic images, a minor systematic background correction was deemed necessary to bring the background levels closer to zero. The correction values for individual filters are shown in Table.~\ref{g7_tab:observation_log}.

The 2022 narrow-band \hb\ and broad-band $g^\prime$ images were processed separately using a newer version of {\sc THELI~(v3)}. We followed the same procedure as for the 2019 data-set with the exception of using a newer Gaia DR3 catalogue\footnote{Recommended setting of the \texttt{ focal plane array handling} parameters were for Gaia DR3 and WFC detector in {\sc THELI (v3)} following: \\ Stability type - exposure; Mosaic type - loose; FPA mode - ignore prior} \citep{Gaia2022}. Resulting combined exposure time is 125 minutes in the narrow-band \hb \ filter and 20 minutes in the broad-band $g^\prime$ filter.

\begin{table*}
    \centering
    \caption{Summary of the broad and narrow-band filters used in this study. The central wavelength, $\lambda$, and the effective width, $\Delta\lambda$, are shown for each filter. We provide information on the average observing conditions and parameters used in the absolute flux calibration of each co-added image. 
    }
    \begin{tabular}{c|lcc|ccc|cc}
        \hline
        {Obs.} & {Filter} & {Filter} & {Filter} & {Average } & {Average } & {Total exp.} & {Systematic bkg.}  & {Zero point}\\
        {year} & name & $\lambda$ [\AA] & $\Delta\lambda$ [\AA] &  seeing [$\arcsec$] & air mass & time [min] &   correction [ct/s]  & calibration  \\
        \hline
        \multirow{5}*{2019} & Sloan $r^\prime$ & 6240 & 1347 & 1.1 & 1.7 & 5 & 1.00 & $24.846\pm0.004$\\ 
        & Sloan $g^\prime$ & 4846 & 1285 & 2.0 & 2.5 & 5 & 0.46  & $24.644\pm0.008$\\ 
        & \hnii & 6568 & 95 & 1.2 & 1.7 & 90 & 0.10  & $21.880\pm0.005$\\ 
        & \hb\ & 4861 & 30 & 2.0 & 1.8 & 15 & 0.014  & $20.705\pm0.017$\\ 
        & \oiii  & 5008 & 100 & 1.9 & 2.2 & 37 & 0.06 & $22.097\pm0.014$\\ 
        & \sii & 6725 & 80 & 2.0 & 2.0 & 30 & 0.14 & $21.840\pm0.011$\\ 
        \hline
        \multirow{2}*{2022} & Sloan $g^\prime$ & 4846 & 1285 & 1.6 & 1.6 & 20 & 0.5 & $26.403\pm0.007$\\ 
        & \hb\ & 4861 & 30 & 1.7 & 1.6 & 125 & -0.02  & $22.115\pm0.015$\\ 
        \hline
    \end{tabular}

    \label{g7_tab:observation_log}
\end{table*}

\begin{figure*}
    \centering
    \includegraphics[width=\linewidth]{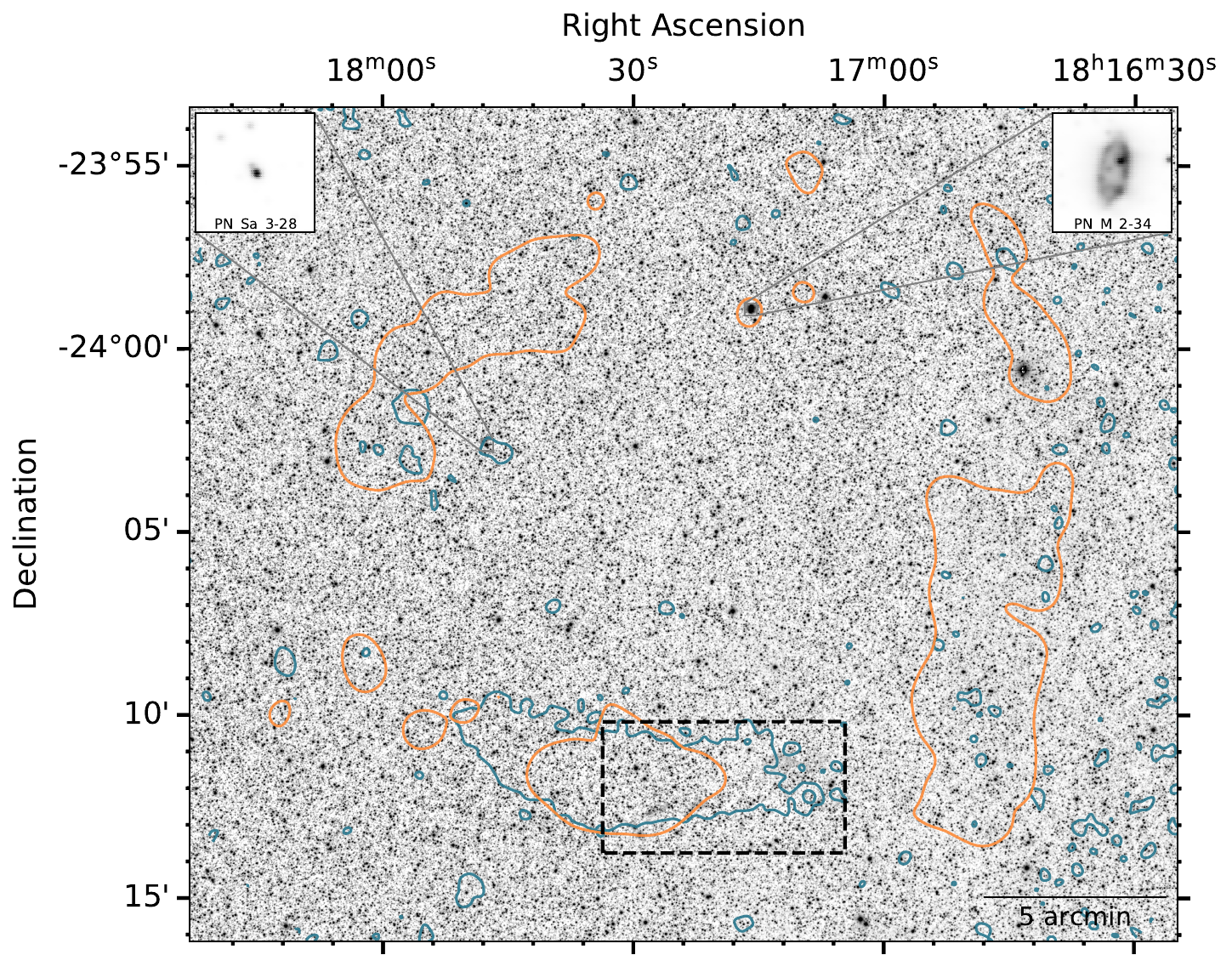}
    \caption{A mosaiced \hnii\ count rate image of SNR G7.7$-$3.7. Orange colour displays 1.4~GHz NVSS radio \citep{Kaplan1998} contours at flux level of 0.003~Jy beam$^{-1}$, while the blue colour represents XMM-Newton (0.5-5.0)~keV contours at flux level of $7\times 10^{-5}$~ct s$^{-1}$ (see figure~1 in \citet{Zhou2018} for full view of radio and X-rays). Dashed rectangle displays a region with the detected optical emission, further used for analysis in Fig.~\ref{g7_fig:diff_all}. The FoV also contains two planetary nebulae (PN~Sa~3-28 \citep{Perek1967} and PN~M~2-34 \citep{Sanduleak1976}) shown in the upper corners of the figure with different scaling and 
    $10\times$ magnification.}
    \label{g7_fig:ha_full}
\end{figure*}

\subsubsection{Flux calibration and continuum subtraction}
We performed a flux calibration to obtain the continuum subtracted images. We used Pan-STARRS DR2 \citep{Flewelling2020} as a calibration catalogue, with stars selected from the $5^{\prime\prime}$ cone centred at the selected shock region (indicated by the rectangle region in Fig.~\ref{g7_fig:ha_full}). Pan-STARRS catalogue brightness range was limited to 13-17~mag in order to avoid star saturation on the brighter end and source confusion on the fainter end. 
The one exception concerns the $g^\prime$ filter data obtained in 2022. Here the range was expanded to 21~mag due to the longer exposure causing saturation of stars in the lower range.

The source catalogue of the observed INT data was constructed using the {\sc SExtractor} \citep{Bertin:1996}. We limited the use of the source catalogue only to stars with a clear detection, represented by the {\sc FLAG=0} condition. An air-mass correction was performed using the average air-mass value per filter listed in Table~\ref{g7_tab:observation_log}. We cross-matched the two catalogues with the distance precision of $1\arcsec$ between two given stars.  
In the case of the broad-band filters, both Pan-STARRS and WFC use identical {\it r-SDSS} and {\it g-SDSS} filters, and after the air-mass correction, both catalogues can be directly compared. 
However, since there are no directly comparable narrow-band filter fluxes, we performed an additional step to calculate them from the available broad-band filter values. 

First, we employed the broad-band photometry of Pan-STARRS to retrieve the effective temperature of selected stars based on the \citet{Castelli2003} stellar atmosphere models (3,500 - 50,000 K). We then found the best temperature model $M(T_{\rm eff})$ and scaling $A$, for every star such that $F(\lambda,T_{\rm eff}) $ = $ A\cdot M(T_{\rm eff})$ using a simple linear interpolation over the grid. We only included stars with fitted temperatures above 3750~K, which we selected in order to avoid extrapolating outside the model grid boundary (3500~K). For every fitted star, we performed synthetic photometry in the wide- and narrow-band filters of the WFC images\footnote{Transmission curves for narrow-band filters were obtained from the WFC website at \url{http://catserver.ing.iac.es/filter/list.php?instrument=WFC}.} to create a star catalogue. 

We obtained the absolute calibration of the image by fitting the ensemble instrumental magnitudes in each filter to determine the global zero-point based on our synthetic model prediction using a linear regression model. We fitted for the zero-point value two times, applying a $3\sigma$-clipping procedure to remove outliers between the two iterations. We also excluded stars brighter than $m<17$, to avoid saturation effects in $r^\prime$ and $i^\prime$ filters. The resulting zero-point values for each WFC filter, which are then applied to the count-rate images, are shown in Table~\ref{g7_tab:observation_log}. 

Finally, to perform the continuum subtraction on the narrow-band images ($r^\prime$ for \hnii\ and \sii; $g^\prime$ for \hb\ and \oiii) we had to further smooth the images to match the point spread functions (PSFs) to common values between images. We have used the source catalogue obtained by {\sc SExtractor} to obtain a distribution of PSFs, and chose a smaller region with non-saturated stars in each image for comparison. We then used the \textit{gaussian\_filter} from the \textit{scipy} library to achieve the best match of PSFs. The following $\sigma$ values for the Gaussian kernel for individual filters were applied: $\sigma $ = $ 0.54$~pix to smooth $r^\prime$ filter for H$\alpha$ correction; $\sigma = 2.25$~pix to smooth $r^\prime$ filter for \sii\ correction; 
$\sigma $ = $ 1.39$~pix to smooth \oiii\ filter to match $g^\prime$ filter. No smoothing was applied for the 2022 \hb \ and $g^\prime$ filters.

After the reprojection, flux calibration, PSF matching and a background adjustment, we were able to obtain the continuum-subtracted images by simple subtraction. We present the resulting images in our region of interest in Fig.~\ref{g7_fig:diff_all}.

\subsection{Spectroscopy}
Low-dispersion long-slit spectra were obtained with the 2.5m INT, on La Palma, Spain on July 4, 2022. Three spectra (1800~s, 900~s, 900~s) of Filament B were taken, resulting in a total exposure time of 3600~s. The R300V grating of the IDS spectrometer was used in conjunction with the Red+2/EEV10 CCD (1$\times$1 binning mode), resulting in a spatial scale of 0.44 arcsec pixel$^{-1}$~covering the range 3500\AA\ -- 9800\AA. The slit has a width of 1\farcs2 and a useful length of 3\farcm3 and it was oriented in the east-west direction. The coordinates of the position centres of each spectrum are given in
Table~\ref{tab:table2}. For the absolute flux calibration the spectrophotometric standard star SP1550+330 was used. The data reduction was performed using standard routines of the {\sc iraf} package \citep{Tody1986}. All spectra were bias-subtracted and flat-field--corrected using a series of well-exposed twilight frames, while  
they were calibrated against the reference spectrum of a CuAr+CuNe arc lamp. The position of the slit, as well as, the spectra are shown in Figs. \ref{g7_fig:image_slit} and \ref{g7_fig:spectra}, respectively.

\begin{table}  
\caption{Log of spectroscopic observations on Filament B. The total exposure time was divided in three individual exposures (see text for details).} \label{tab:table2}
\begin{tabular}{lccccc}  
\noalign{\smallskip}  
\hline  
& \multicolumn{2}{c}{Position centre} & Exp. time & Length$^{\rm a}$\\  
 & $\alpha$ & $\delta$ & (sec) & (arcsec) \\  
\hline  
Pos.1 & 18\h17\m12.7\s & -24\degr11\arcmin05\arcsec & 3600 & 20.2 \\
Pos.2 & 18\h17\m11.6\s & -24\degr11\arcmin05\arcsec & 3600 & 10.6 \\
Pos.3 & 18\h17\m07.5\s & -24\degr11\arcmin05\arcsec & 3600 & 11.4 \\
\hline  
\end{tabular}
\begin{flushleft}
${\rm ^a}$ Extraction aperture lengths for each position.\\
\end{flushleft} 
\end{table}  

\section{Results and discussion}
\label{g7_sec:results_discussion}

\begin{figure*}
    \centering
    \includegraphics[width=0.83\linewidth]{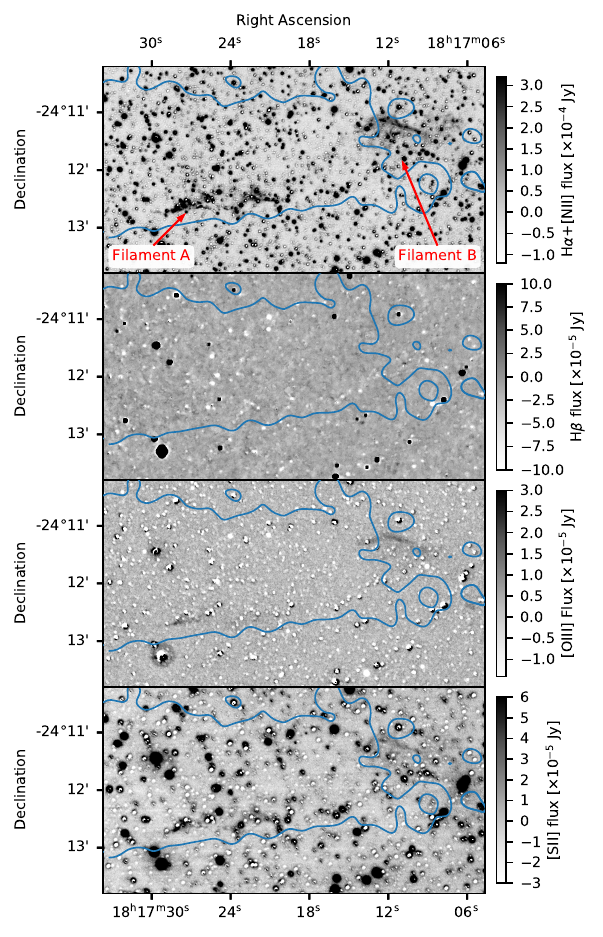}
    \caption{Continuum subtracted images in narrow-band filters from top to bottom - \hnii, \hb, \oiii\ and \sii, displaying the two filaments detected. Contours are the same as in Fig.~\ref{g7_fig:ha_full}.} 

    \label{g7_fig:diff_all}
\end{figure*}

\begin{figure}
    \centering
    \includegraphics[width=1\linewidth]{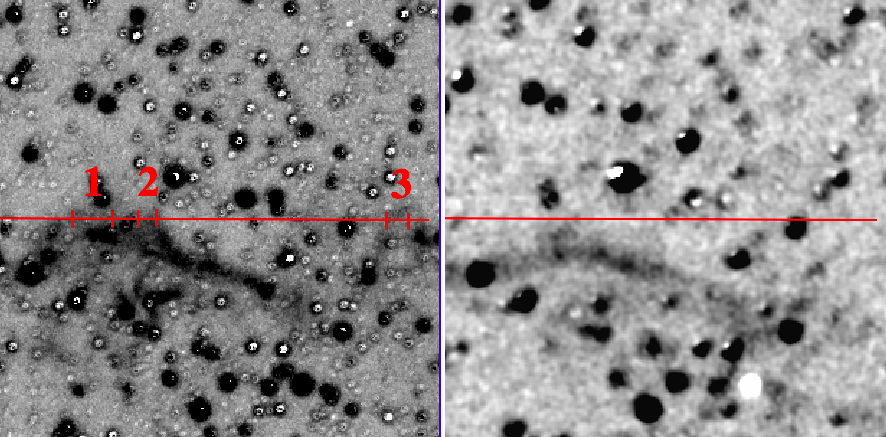}
    \caption{Slit position (\textcolor{red}{red line}) across Filament B region in the \hnii\ (left) and \oiii\ (right) images. The apertures of positions 1, 2 \& 3 are also shown (see Table~\ref{tab:table2}). North is to the top, East to the left. We find a $\sim2$ arcsec offset between the \hnii\ and \oiii\ filaments (see Section~\ref{g7_sec:line_ratio} for further discussion).}
    \label{g7_fig:image_slit}
\end{figure}

\begin{figure}
    \begin{subfigure}{.61\textwidth}
    \includegraphics[width=0.815\linewidth]{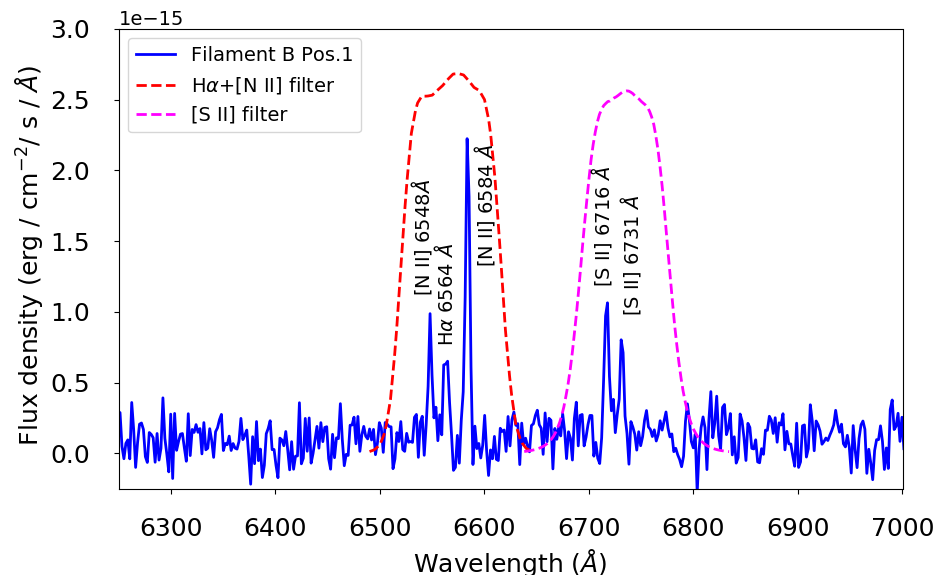}
    \end{subfigure}
    \begin{subfigure}{.61\textwidth}
    \includegraphics[width=0.815\linewidth]{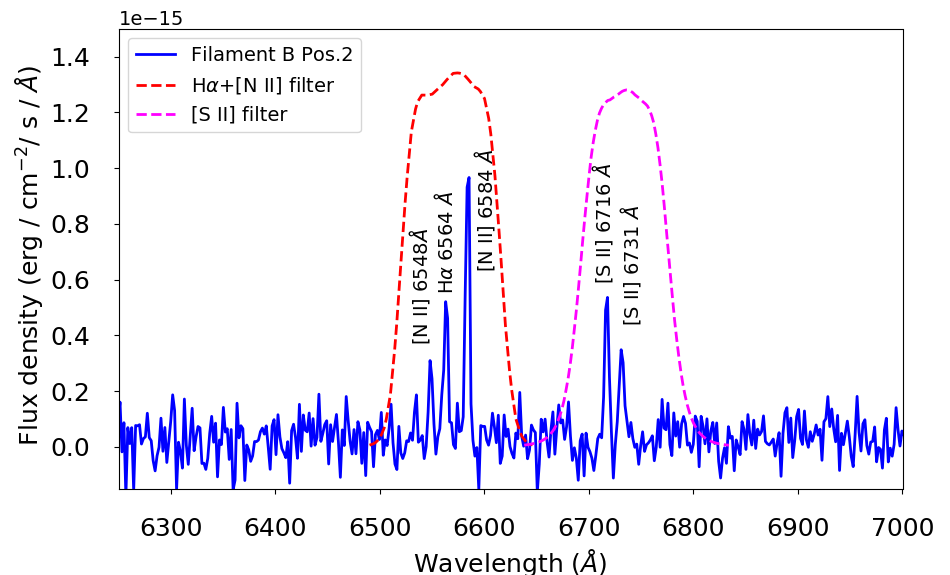}
    \end{subfigure}
    \begin{subfigure}{.61\textwidth}
    \includegraphics[width=0.815\linewidth]{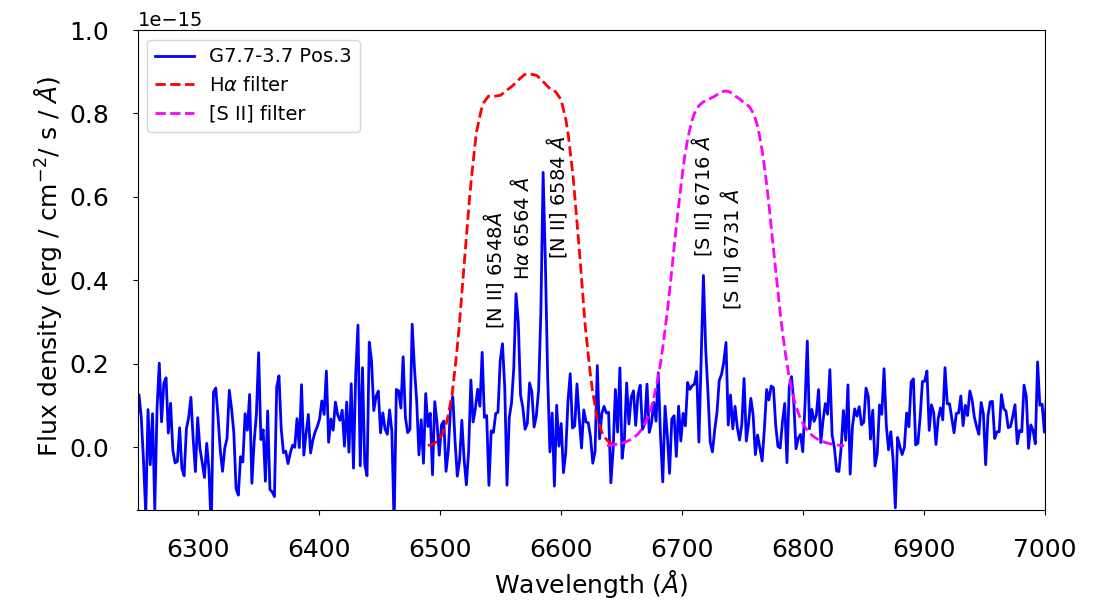}
    \end{subfigure}
    \caption{Long-slit low-resolution spectra for positions 1, 2 and 3 of Filament B (see Table~\ref{table3}) shown in the top, middle and bottom panels, respectively. The WFC narrow filters band-pass (see Table~\ref{g7_tab:observation_log}) are also shown with dash lines (\textcolor{red}{red} for the \hnii\ and \textcolor{magenta}{magenta} for the \sii\ filters).}
    \label{g7_fig:spectra}
\end{figure}

\begin{table}

\caption{Relative line fluxes in different positions on Filament B. The errors of the emission line ratios are calculated through standard
error propagation.}
\begin{tabular}{lllllll}
\hline
\noalign{\smallskip}
 & \multicolumn{2}{c}{Pos.1} & \multicolumn{2}{c}{Pos.2} & \multicolumn{2}{c}{Pos.3} \\ 
  \cline{2-3}  \cline{4-5}  \cline{6-7}
Line (\AA) & F$^{\rm a}$ & S/N$^{\rm b}$ & F & S/N & F & S/N \\
\hline
\nitrogen\ 6548  & 130 & (6)  & 56  & (3)   & 54  & (2) \\
\ha\  6563       & 100 & (9)  & 100 & (4)   & 100 & (3) \\
\nitrogen\ 6584  & 310 & (18) & 159 & (9)   & 150 & (6) \\
\sulfur\ 6716    & 139 & (10) & 97  & (5)   & 96  & (3) \\
\sulfur\ 6731    & 112 & (8)  & 85  & (4)   & 66  & (2) \\
\hline
Absolute \ha\ flux$^{\rm c}$ & \multicolumn{2}{c}{4.1 $\pm$ 0.5} & \multicolumn{2}{c}{3.4 $\pm$ 0.9} & \multicolumn{2}{c}{2.1 $\pm$ 0.5} \\
\sulfur/\ha\  & \multicolumn{2}{c}{2.51 $\pm$ 0.53} & \multicolumn{2}{c}{1.83 $\pm$ 0.71} & \multicolumn{2}{c}{1.63 $\pm$ 1.11} \\
F(6716)/F(6731) & \multicolumn{2}{c}{1.24 $\pm$ 0.22} & \multicolumn{2}{c}{1.14 $\pm$ 0.36} & \multicolumn{2}{c}{1.45 $\pm$ 0.86} \\
\nii/\ha\ & \multicolumn{2}{c}{4.40 $\pm$ 0.95} & \multicolumn{2}{c}{2.15 $\pm$ 0.85} & \multicolumn{2}{c}{2.04 $\pm$ 1.35} \\ 
$n_{\rm e}^{\rm d}$ & \multicolumn{2}{c}{$<400$} & \multicolumn{2}{c}{$<600$} &  \multicolumn{2}{c}{-} \\
\hline
\end{tabular}

${\rm ^a}$ Observed fluxes normalised to F(H$\alpha$)=100 and uncorrected   for interstellar extinction. ${\rm ^b}$ Numbers in parentheses represent the signal-to-noise ratio of the quoted fluxes. $^{\rm c}$ In units of \flux. $^{\rm d}$ In units of $\rm cm^{-3}$. 

\label{table3}
\end{table}

\subsection{Detection of optical emission}
Two faint filamentary structures are detected in \hnii, \oiii\ and marginally in \sii\ narrow-band images, located in the southern part of the SNR (see Fig.~\ref{g7_fig:diff_all}). Similar structures also appear in the SuperCOSMOS H$\alpha$ imaging survey \citep{Parker2005}.
For the purposes of this work, we name these two structures filament A and B, as labeled in Fig.~\ref{g7_fig:diff_all}. Aside from the southern section there is no obvious detection of a diffuse emission in other parts of the SNR.

Both filaments are elongated in the east-west direction. In particular, filament A positioned at the forefront of the X-ray emission (associated to the X-ray forward shock, see lower left region in Fig.~\ref{g7_fig:diff_all}), emitting mainly in \hnii\ and partially in \oiii. On the other hand, Filament B lies in the fainter location in X-rays (upper right in Fig.~\ref{g7_fig:diff_all}) and, although elongated in the east-west direction, it has a more extended diffuse structure compared to Filament A. It is also visible in \hnii\ and \oiii, with a similar, but less significant structure in \sii. {The \hb\ image does not show any significant emission even with the extended 2~hr exposure time.}

\subsection{Line ratio diagnostic of filaments}
\label{g7_sec:line_ratio}

\begin{figure*}
    \centering
    \includegraphics[width=0.85\linewidth]{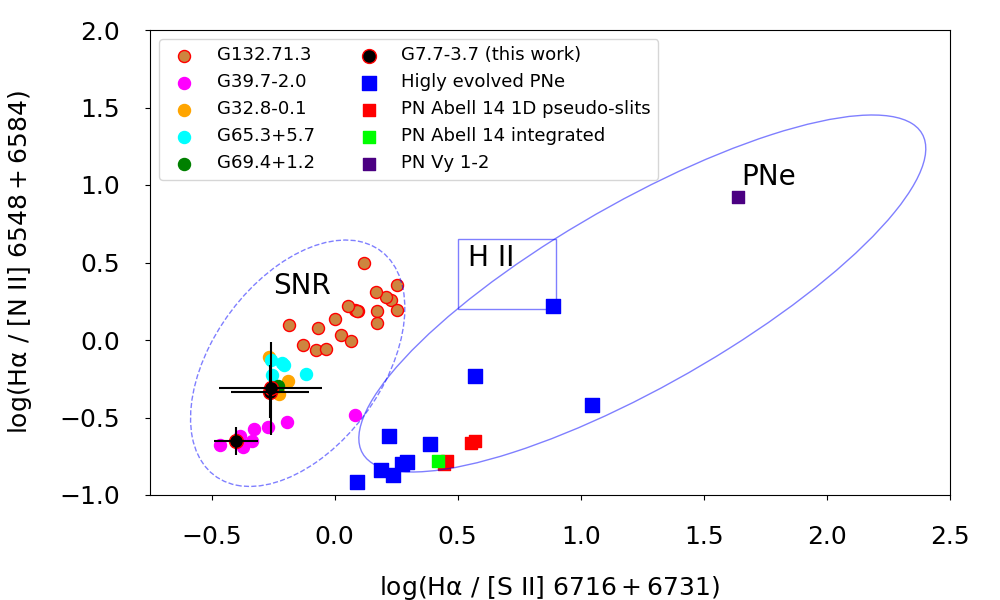}
    \caption{Emission line diagnostic diagram (\citealt{Sab1977}) showing a number of SNRs (\citealt{Boumis2022,Boumis2009,Boumis2007}; \citealt{Mavromatakis2002a,Mavromatakis2002b}) and PNe (PN Vy 1-2 \citep{Akras2015}; highly evolved PNe \citep{Akras2016} and PN Abell~14 \citep{Akras2020} for comparison. The regimes of SNRs, PNe and H~II~regions are also plotted. G7.7-3.7 (black circles) is well-placed in the regime occupied by SNRs with low  H$\alpha$/\sulfur~$\lambda\lambda$6716,6731 and H$\alpha$/\nitrogen~$\lambda\lambda$6548,6584. The line ratios of G7.7-3.7 are taken from Table~\ref{table3}.}
    \label{g7_fig:diagnostic_diagram}
\end{figure*}

The ratio between \sii~doublet lines (at 6716\AA\ and 6731\AA) and H$\alpha$ is commonly used to determine the nature of excitation in the emission regions of SNR and planetary nebulae (PNe). In particular, shock-heated SNR exhibit an enhanced ratio of \sii/H$\alpha > 0.4$ \citep{Long2017}, which is in contrast to the lower ratio typically observed in H~{\sc ii} regions (\sii/H$\alpha \approx 0.1$) or PNe (\sii/H$\alpha < 0.4$). From our imaging alone it is not possible to distinguish the emission mechanism given that narrow-band filters contain several lines from different species (see Fig.~\ref{g7_fig:spectra}), skewing the measurements and thus resulting in ambiguous inferences. 

Thus, to determine the nature of the optical emission in G7.7-3.7, a low-resolution long-slit spectrum of the Filament~B was obtained. Three integrated spectra of Filament~B were extracted with an aperture length of 20.2, 10.6 and 11.4~arcsec (for positions 1, 2 and 3, respectively) and their exact position is shown in Fig.~\ref{g7_fig:image_slit}. The extracted spectra are illustrated in Fig.~\ref{g7_fig:spectra} where the H$\alpha$, \nitrogen~$\lambda\lambda$6548,6584 and \sii~$\lambda\lambda$6716,6731 emission lines are clearly detected. The observed lines fluxes, normalised to F(H$\alpha$)=100, are listed in Table~\ref{table3}. The band-passes of the \hnii\ and \sii\ filters used for the imaging data of G7.7$-$3.7 are also overplotted for reference. 

While Filament~B is clearly detected in our \oxygen~image (see Fig.~\ref{g7_fig:diff_all}), our spectroscopic data do not show any \oxygen\ nor \hb\ line emission (the latter consistent with our null detection in both epochs of deep H$\beta$ imaging). However, the slit position on Filament~B can explain the missing \oxygen~emission lines. Figure~\ref{g7_fig:image_slit} shows the position of the slit (red line) across Filament B in the \hnii\ (left) and \oiii\ (right) images. We can easily discern a spatial offset of $\sim$2~arcsec between the two images. An offset between the \oiii\ and H$\alpha$ regions is expected to be found in SNRs and more generally in shocked gases \citep[e.g.,][]{Boumis2022}. Just behind the shock front, the gas is highly ionised and emission lines like \oiii\ are found. Subsequently, the gas cools down and recombines, resulting in H$\alpha$ emission further back from the shock front. This spatial offset between line emitting regions is thus consistent with the filament moving away from the geometrical centre of the SNR.

In Fig.~\ref{g7_fig:diagnostic_diagram}, we present the common H$\alpha$/\nitrogen~vs.~H$\alpha$/\sii~line ratio diagnostic diagram to disentangle UV-dominated from shock-dominated structures \citep{Sab1977,Akras2022}. This diagram is populated with observed line ratios from a sample of SNRs and PNe together with the regions of parameter space occupied by SNRs, PNe and H~{\sc ii}-regions. The line ratios of G7.7$-$3.7 measured in this work are presented with black circles and are well within the region of SNRs. The H$\alpha$/\sii~line ratio is found to be significantly smaller compared to UV-dominated regions, ranging from $-0.4$ up to $-0.26$ (in logarithmic scale). The low H$\alpha$/\sii~ratio obtained from Filament~B is close to the ones found in another SNR, G39.7$-$7.2 \citep{Boumis2007}. According to our spectroscopic data and the analysis above, the SNR nature of the Filament~B is verified. Based on the observed \sii\ F(6716)/F(6731) ratios and utilising equation 7 of \citet{Samarakoon2018} we can estimate the electron density, $n_{e}$, of both regions. Assuming a characteristic electron temperature of T$_\mathrm{e}$ = 10$^{4}$~K \citep{Samarakoon2018}, we estimated upper limits in Filament~B of $n_e<400$~cm$^{-3}$ in the pos1 and $n_e<600$~cm$^{-3}$ in the pos2. These low electron densities ($n_e<400$~cm$^{-3}$) are consistent with other evolved SNRs, as shown Fig.~\ref{g7_fig:diagnostic_diagram}.

\subsection{On the origin of the optical emission}
G7.7$-$3.7 has been suggested to have its origin in the low-luminosity supernova SN~386~AD \citep{Zhou2018}. However, we did not detect a clear non-radiative Balmer shock emission which, with a few exceptions, predominantly appears in younger SNRs (e.g., SN1006 and Tycho). This type of emission is also more common in SNRs produced by Type Ia SNe, while G7.7$-$3.7 has been proposed to be a result of a low-luminosity Type IIP SN \citep{Zhou2018}. In addition, Balmer-dominated shocks are generally rather faint. Although strong Balmer shock emission can be excluded from the present data, there is still a possibility that some fainter emission remains undetected in the relatively crowded field.

The detected optical emission is consistent with radiative shocks, which can arise from SNRs that have evolved beyond the adiabatic phase, or by collision of the remnant’s forward shock with a dense medium. To assess the plausibility of the first scenario, we estimate the transition radius ($R_{\rm PDS}$) between the Sedov and pressure-driven snowplow (PDS) phase, which marks the evolutionary moment that radiative shocks are expected to be formed \citep{Cioffi1988}:
\begin{equation}\label{eq:RPDS}
{R_{\rm PDS} 
= 14.0 \left( \frac{E}{10^{51}~{\rm erg}} \right)^{\frac{2}{7}} \left( \frac{n_\mathrm{0}}{{\rm cm}^{-3}} \right)^{-\frac{3}{7}} \zeta_m^{-\frac{1}{7}}} \ {\rm pc},
\end{equation} 
where $E$ is the SN energy, $n_\mathrm{0}$ the ambient medium density, and $\zeta_m$ a constant equal to unity for solar metallicity. Adopting an ambient medium density in the range of  $n_\mathrm{0} $ = $ 0.1 - 1~\rm cm^{-3}$ \citep{Zhou2018} and a typical SN energy of $E$ = $10^{51}$~erg the extracted radius is $R_{\rm PDS}$ = $ 14.0 - 37.5$~pc. For the case of a sub-energetic SN explosion of  $E$ = $10^{50}$~erg, as suggested by \citet{Zhou2018}, the corresponding radius is  $R_{\rm PDS}$ = $ 7.3 - 19.5$~pc.  Overall, the extracted transition radius range is larger than the observed radius of G7.$-$3.7, being $R_s \sim 12.8\, d_{4}$, where $d_{4}$ is the distance of the remnants in units of 4~kpc \citep{Milne1986}. This suggests that the remnant is still in the adiabatic phases of SNR evolution in which radiative shocks are not expected.  The only exception is the case of a low energy SN event embedded in a rather high-density ambient medium (for example, for $E$ = $10^{50}$~erg and $n_\mathrm{0} $ = $  1~\rm cm^{-3}$ the transition radius is $R_{\rm PDS}$ = $ 7.3$~pc, smaller than the radius of G7.7$-$3.7). If indeed the remnant has entered the PDS phase then its age should be higher than the Sedov-PDS transition age ($t_{\rm PDS}$, \citealt{Cioffi1988}) i.e.
\begin{equation}
t_{\rm SNR} > t_{\rm PDS} 
= 1.33  \times 10^4 \left( \frac{E}{ 10^{51}~{\rm erg}} \right)^{\frac{3}{14}}\left( \frac{n_\mathrm{0}}{{\rm cm}^{-3}} \right)^{-\frac{4}{7}} \zeta_m^{-\frac{5}{14}} = 8100 \,{\rm yr}.
\end{equation}
Such a scenario is inconsistent with the association of G7.7$-$3.7  with the SN 386~AD, as proposed by \citet{Zhou2018}. However, if this is the case, then radiative shocks should be expected all over the periphery of the SNR---as most evolved SNRs display \citep[e.g.][]{Boumis2009, Fesen1995}-- and not only to the south region of G7.7$-$3.7. 

The problems discussed above can be resolved using an alternative explanation for the optical emission of G7.7$-$3.7, according to which the SNR blast is interacting with a dense shell in the southern region of the remnant. 
Hydrodynamic models of SNRs have shown that the collision of a shock wave with a dense region results in a substantial deceleration of the forward shock's velocity that---depending on the density contrast between the wall and the surrounding medium---can become almost instantaneously radiative \citep{Dwarkadas2005,Dwarkadas2007, Chiotellis2012, Chiotellis2013}. Within the framework of this scenario, the co-spatial X-ray emission of G7.7$-$3.7 can be attributed to the shocked gas that currently has collided with the high-density shell.  Such a scenario has been studied in the context of the extended, young SNR RCW 86, the likely remnant of SN~185 \citep[e.g.][]{Broersen2014}.

Regarding the formation mechanism of the density wall, the spatial locality of the optical emission and its proximity to the explosion centre indicates a circumstellar origin, associated with the mass outflow of the progenitor star. 
The presence of the density wall only at the south of the remnant could be attributed either to a non-spherical symmetric mass outflow from the parent star or/and to ISM density gradients in the form of random dense clouds that where lying close to the progenitor star and they have been mixed with circumstellar material. A third alternative suggests that the progenitor  was a  supersonically moving, mass-losing star  that formed a bow-shaped circumstellar structure around the explosion center. Thus, the emission could be interpreted as the collision of the SNR blast wave with the portion of the bow-shaped circumstellar structure that is lying closest to the explosion centre (i.e., the region around the stagnation point). At the same time, the rest of the remnant still remains within the interior of the bubble being characterised by fast, non-radiative, X-ray emitting shocks.


\subsection{Possible nitrogen overabundance?}

The observed nitrogen enrichment of the shocked gas (see Table~\ref{table3}) advocates for the existence of circumstellar materials near the southern shell of the SNR.
In the optical spectra of most older SNRs, the [NII] emission lines flanking H$\alpha$ emission have a lower flux than the H$\alpha$ emission.
Fig.~\ref{g7_fig:spectra} shows that for G7.7$-$3.7 [N II]/6584 \AA\ dominates over H$\alpha$. Only a few SNRs show this pattern, for example,
radiative shocks in Kepler's SNR \citep[][c.f. knot D3 in Fig. 8]{blair91}, and one filament in Puppis A \citep{sut95}. The relatively high flux of [NII] is in those cases attributed to an enhanced nitrogen abundance, taken to be caused by pre-supernova mass loss.
The gas in the shell can also be a mixture of circumstellar and interstellar materials. The X-ray-emitting gas in the southern shell of G7.7$-$3.7 shows subsolar abundances from Oxygen to Iron \citep{Zhou2018}. Combined with the optical observation, this suggests that the pre-supernova wind materials are rich in Nitrogen but not in heavier elements such as Oxygen.

Nitrogen-rich mass outflows in the form of stellar winds are expected by red supergiant (RSGs) progenitor stars \citep[e.g.][]{Origlia2016} or, in the low mass regime, by stars in the asymptotic giant branch (AGB) with initial mass $ \ge 4~\rm M_{\odot}$ \citep[e.g.][]{Karakas2018}. Nitrogen-rich Wolf-Rayet stars \citep{Crowther2007} is a third alternative but it is less likely as their stellar winds excavate very extended circumstellar structures due to their high wind mechanical luminosity \citep{Smith2014}.

If indeed the SNR is evolving within the wind bubble of its progenitor star, we can provide a draft estimation of its age by employing the self-similar SNR evolution models of \citet{Chevalier1982}. In particular, describing the SN ejecta density  with a power law of $\rho_{ej} \propto r^{-n}$, with n=7 and the wind blown bubble in which the remnant is evolving with a density profile of $\rho_{AM}=q~r^{-s}$, where s=2  and $q= \frac { \dot{M_w}} { 4 \pi u_w }$, with  $\dot{M_w}$ and $u_w$, the stellar wind mass loss rate and terminal velocity respectively, the age of the remnant is given by:

\begin{equation}\label{eq:Rs_Chev}
{\rm t_{snr}= \left[ \frac{R_{s}} {1.3} \times \left( \frac{A~g^n}{q} \right)^{\frac{1}{s-n}} \right]^{\frac{n-s}{n-3}}}.
\end{equation}
where A= 0.27,  $g= \left[ \left( \frac{25}{21 \pi}\right) \left(\frac{E_{ej}^2}{M_{ej}}\right) \right]^\frac{1}{7}$  with $E_{ej}$ and $M_{ej}$ the SN energy and ejecta mass, respectively and $R_s= 12.8~d_4$~pc, the current radius of the remnant.  Adopting typical values for a Type II SN resulted by a red supergiant progenitor, namely $M_{ej} = 6-9~\rm M_{\odot}$, $E_{ej}= 10^{51}$~erg,   $\dot{M_w}= 10^{-6}-10^{-5}~\rm M_{\odot}yr^{-1}$ and $u_w= 50-100~\rm km~s^{-1}$ we extract a range for the SNR age equal to $t_{snr}= (1120-2610)d_4^{5/7}$~yrs. For the case of a Type Ia progenitor, employing to Eq. \ref{eq:Rs_Chev} the relevant SN and AGB wind properties:  $M_{ej} = 1-1.4~\rm M_{\odot}$, $E_{ej}= 10^{51}$~erg,   $\dot{M_w}= 10^{-6}-10^{-5}~\rm M_{\odot}yr^{-1}$ and $u_w= 5-20~\rm km~s^{-1}$, the corresponding range of the SNR age is  $t_{snr}= (1070-2920)d_4^{5/7}$~yrs. Both scenarios do not contradict to a possible association of G7.7-3.7 with the historical SN 386~AD as in such a case the current SNR age had to be $t_{snr}= 1637$~yrs. Thus, a possible association of G7.7-3.7 with SN 386 seems
to be alighted to an evolutionary scenario according to which
the remnant is evolving within the wind bubble formed by a
Type II or Type Ia stellar progenitor and currently is partially
colliding with a local circumstellar shell at the southern region
of the SNR. Nevertheless, detailed hydrodynamic simulations are required to assess this statement. 

\section{Conclusions}

We performed the first deep optical study in the direction of the SNR G7.7$-$3.7. The region of the SNR is mostly clear of diffuse optical emission. The only exception is in the south of the remnant, where the radio and X-ray emission has been previously detected. Here, we detected two filamentary structures clearly visible in the \hnii filter, with fainter emission in the forbidden-line filters \oiii\ and \sii. Both filaments -- in particular Filament A -- appear to correspond well with the location of the forward shock seen in X-rays.

Follow-up spectroscopic observations of one of the filaments revealed large ratios of \sii/\ha \ =$ (1.6-2.5)$, expected for SNR related shocks. In addition, the \sii\ doublet ratio also suggests upper limits of electron density of the gas (<400 and <600 cm$^{-3}$, for positions 1 \& 2 respectively; see Table~\ref{table3} for more details). The emission line diagnostics suggest similar ratios as observed in other SNR, providing robust evidence against an HII region origin. Based on the filamentary morphology, spatial coincidence with the X-ray emission, and spectroscopic diagnostic line ratios that the optical emission discovered is associated to the SNR  G7.7$-$3.7 originated by radiative shocks.

Our calculations show that the optical emission from the SNR is more likely to originate from the collision of forward shock with the locally higher density ISM. The SNR evolutionary phase could still be consistent with the radiative post-adiabatic phase, but this requires a sub-energetic explosion ($E$ = $10^{50}$~erg) that occurred in an environment with $n_\mathrm{0}$ = $1$~cm$^{-3}$. This explanation would, however, put the age of the remnant closer to 8000~yrs, which is inconsistent with the ionisation age of the SNR as determined from X-ray observations. 
Such an inconsistency can be resolved if the optically emitting regions are associated with patches of gas with higher densities. Several young SNRs indeed show such radiative shocks like the historical SNR RCW 86 (SN 185), which is also a quite extended SNR, and Kepler's SNR (SN 1604), a more compact SNR. Moreover, the large ratio of [NII] over H$\alpha$ emission suggest that the gas is enriched in nitrogen, suggesting that the shock moves through material lost by the progenitor star. This is by itself an indication that G7.7$-$3.7 is a young or moderately aged SNR.

Our optical study does not allow to make a firm conclusion about whether G7.7$-$3.7 is indeed the remnant of the historical SN 386~AD. Despite the presence of radiative shocks, normally an indication of an older SNR, there are counter examples of young SNRs with radiative shocks, and the spectral/morphological properties of the optical emission from G7.7$-$3.7 matches these counter example. Given the interest in identifying new historical SNRs, our study suggest it is worthwhile to continue investigating G7.7$-$3.7 as the potential remnant of SN 386~AD.

\section*{Acknowledgements}
The work of VD is supported by a grant from NWO graduate program/GRAPPA-PhD program. VD also acknowledges support from the LKBF, subsidy no. 19.2.027. JVHS acknowledges support from STFC grant ST/R000824/1. PZ acknowledges the support from the NWO Veni Fellowship, grant no. 639.041.647 and NSFC grant 11590781.
SA thanks the support under the grant 5077 financed by IAASARS/NOA. AdB thanks the support from the Spanish Government Ministerio de Ciencia e Innovaci\'on through grants PGC-2018-091,3741-B-C22 and SEV 2015-0548, and from the Canarian Agency for Research, Innovation and Information Society (ACIISI), of the Canary Islands Government, and the European Regional Development Fund (ERDF), under grant with reference ProID2017010115.
We would like to acknowledge Jiang-Tao Li for the initial exploratory optical images of the field. VD and JVHS would also like to acknowledge the support of Leo Hern\'andez-Su\'arez. AC gratefully acknowledges Mano Trampouli for the fruitful discussions and inspiration as well as the students of the 4rth Lykeion Acharnon for their support and motivation. \\
This research made use of {\sc astropy}, a community-developed core {\sc python} package for Astronomy \citep{AstropyCollaboration:2013}, {\sc matplotlib} \citep{Hunter:2007}, Numpy \citep{Harris2020}, SciPy \citep{Virtanen2020}, APLpy \citep{Robitaille2019}, Pandas \citep{reback2020pandas} and Lmfit \citep{Newville2014}. We further made use of {\sc SAOImage DS9} \citep{Joye2003} and the SAO/NASA Astrophysics Data System. 
INT-WFC photometry was obtained as part of the ING.NL.19B.006 programme.
The INT is operated on the island of La Palma by the Isaac Newton Group of Telescopes in the Spanish Observatorio del Roque de los Muchachos of the Instituto de Astrof\'isica de Canarias.
The Pan-STARRS1 Surveys (PS1) and the PS1 public science archive have been made possible through contributions by the Institute for Astronomy, the University of Hawaii, the Pan-STARRS Project Office, the Max-Planck Society and its participating institutes, the Max Planck Institute for Astronomy, Heidelberg and the Max Planck Institute for Extraterrestrial Physics, Garching, The Johns Hopkins University, Durham University, the University of Edinburgh, the Queen's University Belfast, the Harvard-Smithsonian Center for Astrophysics, the Las Cumbres Observatory Global Telescope Network Incorporated, the National Central University of Taiwan, the Space Telescope Science Institute, the National Aeronautics and Space Administration under Grant No. NNX08AR22G issued through the Planetary Science Division of the NASA Science Mission Directorate, the National Science Foundation Grant No. AST-1238877, the University of Maryland, Eotvos Lorand University (ELTE), the Los Alamos National Laboratory, and the Gordon and Betty Moore Foundation. This work has made use of data from the European Space Agency (ESA) mission {\it Gaia} (\url{https://www.cosmos.esa.int/gaia}), processed by the {\it Gaia} Data Processing and Analysis Consortium (DPAC, \url{https://www.cosmos.esa.int/web/gaia/dpac/consortium}). Funding for the DPAC has been provided by national institutions, in particular the institutions participating in the {\it Gaia} Multilateral Agreement. We would also like to thank the anonymous referee for their comments that greatly improved this paper.

\section*{Data Availability}
The data underlying this article will be available at Zenodo repository after the completion of the referee process (\url{https://doi.org/10.5281/zenodo.5121367}).



\bibliographystyle{mnras}
\bibliography{bibliography.bib} 

\begin{thebibliography}{}
\makeatletter
\relax
\def\mn@urlcharsother{\let\do\@makeother \do\$\do\&\do\#\do\^\do\_\do\%\do\~}
\def\mn@doi{\begingroup\mn@urlcharsother \@ifnextchar [ {\mn@doi@}
  {\mn@doi@[]}}
\def\mn@doi@[#1]#2{\def\@tempa{#1}\ifx\@tempa\@empty \href
  {http://dx.doi.org/#2} {doi:#2}\else \href {http://dx.doi.org/#2} {#1}\fi
  \endgroup}
\def\mn@eprint#1#2{\mn@eprint@#1:#2::\@nil}
\def\mn@eprint@arXiv#1{\href {http://arxiv.org/abs/#1} {{\tt arXiv:#1}}}
\def\mn@eprint@dblp#1{\href {http://dblp.uni-trier.de/rec/bibtex/#1.xml}
  {dblp:#1}}
\def\mn@eprint@#1:#2:#3:#4\@nil{\def\@tempa {#1}\def\@tempb {#2}\def\@tempc
  {#3}\ifx \@tempc \@empty \let \@tempc \@tempb \let \@tempb \@tempa \fi \ifx
  \@tempb \@empty \def\@tempb {arXiv}\fi \@ifundefined
  {mn@eprint@\@tempb}{\@tempb:\@tempc}{\expandafter \expandafter \csname
  mn@eprint@\@tempb\endcsname \expandafter{\@tempc}}}

\bibitem[\protect\citeauthoryear{Acero et~al.,}{Acero et~al.}{2016}]{Acero2016}
Acero F.,  et~al., 2016, \mn@doi [The Astrophysical Journal Supplement Series]
  {10.3847/0067-0049/224/1/8}, 224, 8

\bibitem[\protect\citeauthoryear{{Akras}, {Boumis}, {Meaburn}, {Alikakos},
  {L{\'o}pez}  \& {Gon{\c{c}}alves}}{{Akras} et~al.}{2015}]{Akras2015}
{Akras} S.,  {Boumis} P.,  {Meaburn} J.,  {Alikakos} J.,  {L{\'o}pez} J.~A.,
  {Gon{\c{c}}alves} D.~R.,  2015, \mn@doi [\mnras] {10.1093/mnras/stv1468},
  \href {https://ui.adsabs.harvard.edu/abs/2015MNRAS.452.2911A} {452, 2911}

\bibitem[\protect\citeauthoryear{{Akras}, {Clyne}, {Boumis}, {Monteiro},
  {Gon{\c{c}}alves}, {Redman}  \& {Williams}}{{Akras} et~al.}{2016}]{Akras2016}
{Akras} S.,  {Clyne} N.,  {Boumis} P.,  {Monteiro} H.,  {Gon{\c{c}}alves}
  D.~R.,  {Redman} M.~P.,   {Williams} S.,  2016, \mn@doi [\mnras]
  {10.1093/mnras/stw038}, \href
  {https://ui.adsabs.harvard.edu/abs/2016MNRAS.457.3409A} {457, 3409}

\bibitem[\protect\citeauthoryear{{Akras}, {Monteiro}, {Aleman}, {Farias}, {May}
   \& {Pereira}}{{Akras} et~al.}{2020}]{Akras2020}
{Akras} S.,  {Monteiro} H.,  {Aleman} I.,  {Farias} M. A.~F.,  {May} D.,
  {Pereira} C.~B.,  2020, \mn@doi [\mnras] {10.1093/mnras/staa383}, \href
  {https://ui.adsabs.harvard.edu/abs/2020MNRAS.493.2238A} {493, 2238}

\bibitem[\protect\citeauthoryear{{Akras} et~al.,}{{Akras}
  et~al.}{2022}]{Akras2022}
{Akras} S.,  et~al., 2022, \mn@doi [\mnras] {10.1093/mnras/stac632}, \href
  {https://ui.adsabs.harvard.edu/abs/2022MNRAS.512.2202A} {512, 2202}

\bibitem[\protect\citeauthoryear{Arendt}{Arendt}{1989}]{Arendt1989}
Arendt R.~G.,  1989, \mn@doi [The Astrophysical Journal Supplement Series]
  {10.1086/191337}, 70, 181

\bibitem[\protect\citeauthoryear{{Astropy Collaboration} et~al.,}{{Astropy
  Collaboration} et~al.}{2013}]{AstropyCollaboration:2013}
{Astropy Collaboration} et~al., 2013, \mn@doi [\aap]
  {10.1051/0004-6361/201322068}, \href
  {https://ui.adsabs.harvard.edu/abs/2013A&A...558A..33A} {558, A33}

\bibitem[\protect\citeauthoryear{{Bertin} \& {Arnouts}}{{Bertin} \&
  {Arnouts}}{1996}]{Bertin:1996}
{Bertin} E.,  {Arnouts} S.,  1996, \mn@doi [\aaps] {10.1051/aas:1996164}, \href
  {https://ui.adsabs.harvard.edu/abs/1996A&AS..117..393B} {117, 393}

\bibitem[\protect\citeauthoryear{Bilikova, Williams, Chu, Gruendl  \&
  Lundgren}{Bilikova et~al.}{2007}]{Bilikova2007}
Bilikova J.,  Williams R. N.~M.,  Chu Y.-H.,  Gruendl R.~A.,   Lundgren B.~F.,
  2007, \mn@doi [The Astronomical Journal] {10.1086/522302}, 134, 2308

\bibitem[\protect\citeauthoryear{{Blair}, {Long}  \& {Vancura}}{{Blair}
  et~al.}{1991}]{blair91}
{Blair} W.~P.,  {Long} K.~S.,   {Vancura} O.,  1991, \mn@doi [\apj]
  {10.1086/169583}, \href
  {https://ui.adsabs.harvard.edu/abs/1991ApJ...366..484B} {366, 484}

\bibitem[\protect\citeauthoryear{{Boumis} et~al.,}{{Boumis}
  et~al.}{2007}]{Boumis2007}
{Boumis} P.,  et~al., 2007, \mn@doi [\mnras]
  {10.1111/j.1365-2966.2007.12276.x}, \href
  {https://ui.adsabs.harvard.edu/abs/2007MNRAS.381..308B} {381, 308}

\bibitem[\protect\citeauthoryear{{Boumis}, {Xilouris}, {Alikakos},
  {Christopoulou}, {Mavromatakis}, {Katsiyannis}  \& {Goudis}}{{Boumis}
  et~al.}{2009}]{Boumis2009}
{Boumis} P.,  {Xilouris} E.~M.,  {Alikakos} J.,  {Christopoulou} P.~E.,
  {Mavromatakis} F.,  {Katsiyannis} A.~C.,   {Goudis} C.~D.,  2009, \mn@doi
  [\aap] {10.1051/0004-6361/200811474}, \href
  {https://ui.adsabs.harvard.edu/abs/2009A&A...499..789B} {499, 789}

\bibitem[\protect\citeauthoryear{{Boumis} et~al.,}{{Boumis}
  et~al.}{2022}]{Boumis2022}
{Boumis} P.,  et~al., 2022, \mn@doi [\mnras] {10.1093/mnras/stac412}, \href
  {https://ui.adsabs.harvard.edu/abs/2022MNRAS.512.1658B} {512, 1658}

\bibitem[\protect\citeauthoryear{{Broersen}, {Chiotellis}, {Vink}  \&
  {Bamba}}{{Broersen} et~al.}{2014}]{Broersen2014}
{Broersen} S.,  {Chiotellis} A.,  {Vink} J.,   {Bamba} A.,  2014, \mn@doi
  [\mnras] {10.1093/mnras/stu667}, \href
  {https://ui.adsabs.harvard.edu/abs/2014MNRAS.441.3040B} {441, 3040}

\bibitem[\protect\citeauthoryear{{Castelli} \& {Kurucz}}{{Castelli} \&
  {Kurucz}}{2003}]{Castelli2003}
{Castelli} F.,  {Kurucz} R.~L.,  2003, in {Piskunov} N.,  {Weiss} W.~W.,
  {Gray} D.~F.,  eds,  Proceedings of the IAU Symp. No 210 Vol. 210, Modelling
  of Stellar Atmospheres. p.~A20 (\mn@eprint {arXiv} {astro-ph/0405087})

\bibitem[\protect\citeauthoryear{{Chevalier}}{{Chevalier}}{1982}]{Chevalier1982}
{Chevalier} R.~A.,  1982, \mn@doi [\apj] {10.1086/160126}, \href
  {https://ui.adsabs.harvard.edu/abs/1982ApJ...258..790C} {258, 790}

\bibitem[\protect\citeauthoryear{Chevalier, Raymond  \& Kirshner}{Chevalier
  et~al.}{1980}]{Chevalier1980}
Chevalier R.~A.,  Raymond J.~C.,   Kirshner R.~P.,  1980, \mn@doi [The
  Astrophysical Journal] {10.1086/157623}, 235, 186

\bibitem[\protect\citeauthoryear{{Chiotellis}, {Schure}  \&
  {Vink}}{{Chiotellis} et~al.}{2012}]{Chiotellis2012}
{Chiotellis} A.,  {Schure} K.~M.,   {Vink} J.,  2012, \mn@doi [\aap]
  {10.1051/0004-6361/201014754}, \href
  {https://ui.adsabs.harvard.edu/abs/2012A&A...537A.139C} {537, A139}

\bibitem[\protect\citeauthoryear{{Chiotellis}, {Kosenko}, {Schure}, {Vink}  \&
  {Kaastra}}{{Chiotellis} et~al.}{2013}]{Chiotellis2013}
{Chiotellis} A.,  {Kosenko} D.,  {Schure} K.~M.,  {Vink} J.,   {Kaastra} J.~S.,
   2013, \mn@doi [\mnras] {10.1093/mnras/stt1406}, \href
  {https://ui.adsabs.harvard.edu/abs/2013MNRAS.435.1659C} {435, 1659}

\bibitem[\protect\citeauthoryear{{Cioffi}, {McKee}  \& {Bertschinger}}{{Cioffi}
  et~al.}{1988}]{Cioffi1988}
{Cioffi} D.~F.,  {McKee} C.~F.,   {Bertschinger} E.,  1988, \mn@doi [\apj]
  {10.1086/166834}, \href
  {https://ui.adsabs.harvard.edu/abs/1988ApJ...334..252C} {334, 252}

\bibitem[\protect\citeauthoryear{Crowther}{Crowther}{2007}]{Crowther2007}
Crowther P.~A.,  2007, \mn@doi [Annual Review of Astronomy and Astrophysics]
  {10.1146/annurev.astro.45.051806.110615}, 45, 177

\bibitem[\protect\citeauthoryear{Dubner, Giacani, Goss, Moffett  \&
  Holdaway}{Dubner et~al.}{1996}]{Dubner1996}
Dubner G.~M.,  Giacani E.~B.,  Goss W.~M.,  Moffett D.~A.,   Holdaway M.,
  1996, \mn@doi [The Astronomical Journal] {10.1086/117875}, 111, 1304

\bibitem[\protect\citeauthoryear{{Dwarkadas}}{{Dwarkadas}}{2005}]{Dwarkadas2005}
{Dwarkadas} V.~V.,  2005, \mn@doi [\apj] {10.1086/432109}, \href
  {https://ui.adsabs.harvard.edu/abs/2005ApJ...630..892D} {630, 892}

\bibitem[\protect\citeauthoryear{{Dwarkadas}}{{Dwarkadas}}{2007}]{Dwarkadas2007}
{Dwarkadas} V.~V.,  2007, \mn@doi [\apj] {10.1086/520670}, \href
  {https://ui.adsabs.harvard.edu/abs/2007ApJ...667..226D} {667, 226}

\bibitem[\protect\citeauthoryear{{Erben} et~al.,}{{Erben}
  et~al.}{2005}]{Erben2005}
{Erben} T.,  et~al., 2005, \mn@doi [Astronomische Nachrichten]
  {10.1002/asna.200510396}, \href
  {https://ui.adsabs.harvard.edu/abs/2005AN....326..432E} {326, 432}

\bibitem[\protect\citeauthoryear{{Fesen}, {Downes}, {Wallace}  \&
  {Normandeau}}{{Fesen} et~al.}{1995}]{Fesen1995}
{Fesen} R.~A.,  {Downes} R.~A.,  {Wallace} D.,   {Normandeau} M.,  1995,
  \mn@doi [\aj] {10.1086/117736}, \href
  {https://ui.adsabs.harvard.edu/abs/1995AJ....110.2876F} {110, 2876}

\bibitem[\protect\citeauthoryear{{Flewelling} et~al.,}{{Flewelling}
  et~al.}{2020}]{Flewelling2020}
{Flewelling} H.~A.,  et~al., 2020, \mn@doi [\apjs] {10.3847/1538-4365/abb82d},
  \href {https://ui.adsabs.harvard.edu/abs/2020ApJS..251....7F} {251, 7}

\bibitem[\protect\citeauthoryear{{Gaia Collaboration} et~al.,}{{Gaia
  Collaboration} et~al.}{2016a}]{Gaia2016a}
{Gaia Collaboration} et~al., 2016a, \mn@doi [\aap]
  {10.1051/0004-6361/201629272}, \href
  {https://ui.adsabs.harvard.edu/abs/2016A&A...595A...1G} {595, A1}

\bibitem[\protect\citeauthoryear{{Gaia Collaboration} et~al.,}{{Gaia
  Collaboration} et~al.}{2016b}]{Gaia2016b}
{Gaia Collaboration} et~al., 2016b, \mn@doi [\aap]
  {10.1051/0004-6361/201629512}, \href
  {https://ui.adsabs.harvard.edu/abs/2016A&A...595A...2G} {595, A2}

\bibitem[\protect\citeauthoryear{{Gaia Collaboration} et~al.,}{{Gaia
  Collaboration} et~al.}{2022}]{Gaia2022}
{Gaia Collaboration} et~al., 2022, arXiv e-prints, \href
  {https://ui.adsabs.harvard.edu/abs/2022arXiv220800211G} {p. arXiv:2208.00211}

\bibitem[\protect\citeauthoryear{Gardner, Whiteoak  \& Morris}{Gardner
  et~al.}{1969}]{Gardner1969}
Gardner F.,  Whiteoak J.,   Morris D.,  1969, \mn@doi [Australian Journal of
  Physics] {10.1071/PH690821}, 22, 821

\bibitem[\protect\citeauthoryear{{Ghavamian}, {Raymond}, {Hartigan}  \&
  {Blair}}{{Ghavamian} et~al.}{2000}]{Ghavamian2000}
{Ghavamian} P.,  {Raymond} J.,  {Hartigan} P.,   {Blair} W.~P.,  2000, \mn@doi
  [\apj] {10.1086/308811}, \href
  {https://ui.adsabs.harvard.edu/abs/2000ApJ...535..266G} {535, 266}

\bibitem[\protect\citeauthoryear{{Giacani}, {Loiseau}, {Smith}, {Dubner}  \&
  {Iacobelli}}{{Giacani} et~al.}{2010}]{Giacani2010}
{Giacani} E.,  {Loiseau} N.,  {Smith} M.~J.~S.,  {Dubner} G.,   {Iacobelli} M.,
   2010, in {Comastri} A.,  {Angelini} L.,   {Cappi} M.,  eds,  American
  Institute of Physics Conference Series Vol. 1248, X-ray Astronomy 2009;
  Present Status, Multi-Wavelength Approach and Future Perspectives. pp 39--40,
  \mn@doi{10.1063/1.3475267}

\bibitem[\protect\citeauthoryear{Green}{Green}{2015}]{Green2015}
Green D.~A.,  2015, \mn@doi [Monthly Notices of the Royal Astronomical Society]
  {10.1093/mnras/stv1885}, 454, 1517

\bibitem[\protect\citeauthoryear{Harris et~al.,}{Harris
  et~al.}{2020}]{Harris2020}
Harris C.~R.,  et~al., 2020, \mn@doi [Nature] {10.1038/s41586-020-2649-2}, 585,
  357

\bibitem[\protect\citeauthoryear{Heng}{Heng}{2010}]{Heng2010}
Heng K.,  2010, \mn@doi [Publications of the Astronomical Society of Australia]
  {10.1071/AS09057}, 27, 23

\bibitem[\protect\citeauthoryear{{Hester}, {Raymond}  \& {Blair}}{{Hester}
  et~al.}{1994}]{Hester1994}
{Hester} J.~J.,  {Raymond} J.~C.,   {Blair} W.~P.,  1994, \mn@doi [\apj]
  {10.1086/173598}, \href
  {https://ui.adsabs.harvard.edu/abs/1994ApJ...420..721H} {420, 721}

\bibitem[\protect\citeauthoryear{{Hunter, J. D.}}{{Hunter, J.
  D.}}{2007}]{Hunter:2007}
{Hunter, J. D.} 2007, {Computing In Science \& Engineering}, 9, 90

\bibitem[\protect\citeauthoryear{Joye \& Mandel}{Joye \&
  Mandel}{2003}]{Joye2003}
Joye W.,  Mandel E.,  2003, in Payne H.,  Jedrzejewski R.,   Hook R.,  eds,
  Astronomical Society of the Pacific Conference Series Vol. 295, Astronomical
  Data Analysis Software and Systems XII. p.~489

\bibitem[\protect\citeauthoryear{Kaplan, Condon, Arzoumanian  \& Cordes}{Kaplan
  et~al.}{1998}]{Kaplan1998}
Kaplan D.~L.,  Condon J.~J.,  Arzoumanian Z.,   Cordes J.~M.,  1998, \mn@doi
  [The Astrophysical Journal Supplement Series] {10.1086/313153}, 119, 75

\bibitem[\protect\citeauthoryear{Karakas, Lugaro, Carlos, Cseh, Kamath  \&
  García-Hernández}{Karakas et~al.}{2018}]{Karakas2018}
Karakas A.~I.,  Lugaro M.,  Carlos M.,  Cseh B.,  Kamath D.,
  García-Hernández D.~A.,  2018, \mn@doi [Monthly Notices of the Royal
  Astronomical Society] {10.1093/mnras/sty625}, 477, 421

\bibitem[\protect\citeauthoryear{{Leonidaki}, {Boumis}  \& {Zezas}}{{Leonidaki}
  et~al.}{2013}]{Leonidaki2013}
{Leonidaki} I.,  {Boumis} P.,   {Zezas} A.,  2013, \mn@doi [\mnras]
  {10.1093/mnras/sts324}, \href
  {https://ui.adsabs.harvard.edu/abs/2013MNRAS.429..189L} {429, 189}

\bibitem[\protect\citeauthoryear{Long}{Long}{2017}]{Long2017}
Long K.~S.,  2017, in , Handbook of Supernovae.
Springer International Publishing, Cham, pp 2005--2040 (\mn@eprint {arXiv}
  {1712.05331}), \mn@doi{10.1007/978-3-319-21846-5_90}, \url
  {http://link.springer.com/10.1007/978-3-319-21846-5{\_}90}

\bibitem[\protect\citeauthoryear{{Mavromatakis}, {Boumis}  \&
  {Paleologou}}{{Mavromatakis} et~al.}{2002a}]{Mavromatakis2002a}
{Mavromatakis} F.,  {Boumis} P.,   {Paleologou} E.~V.,  2002a, \mn@doi [\aap]
  {10.1051/0004-6361:20020398}, \href
  {https://ui.adsabs.harvard.edu/abs/2002A&A...387..635M} {387, 635}

\bibitem[\protect\citeauthoryear{{Mavromatakis}, {Boumis}, {Papamastorakis}  \&
  {Ventura}}{{Mavromatakis} et~al.}{2002b}]{Mavromatakis2002b}
{Mavromatakis} F.,  {Boumis} P.,  {Papamastorakis} J.,   {Ventura} J.,  2002b,
  \mn@doi [\aap] {10.1051/0004-6361:20020511}, \href
  {https://ui.adsabs.harvard.edu/abs/2002A&A...388..355M} {388, 355}

\bibitem[\protect\citeauthoryear{Mccray}{Mccray}{2016}]{Mccray2016}
Mccray R.,  2016, {Handbook of Supernovae}.
Springer International Publishing, Cham, \mn@doi{10.1007/978-3-319-20794-0},
  \url {http://link.springer.com/10.1007/978-3-319-20794-0}

\bibitem[\protect\citeauthoryear{Milisavljevic \& Fesen}{Milisavljevic \&
  Fesen}{2013}]{Milisavljevic2013}
Milisavljevic D.,  Fesen R.~A.,  2013, \mn@doi [The Astrophysical Journal]
  {10.1088/0004-637X/772/2/134}, 772, 134

\bibitem[\protect\citeauthoryear{Milne \& Dickel}{Milne \&
  Dickel}{1974}]{Milne1974}
Milne D.,  Dickel J.,  1974, \mn@doi [Australian Journal of Physics]
  {10.1071/PH740549}, 27, 549

\bibitem[\protect\citeauthoryear{Milne, Roger, Kesteven, Haynes, Wellington  \&
  Stewart}{Milne et~al.}{1986}]{Milne1986}
Milne D.~K.,  Roger R.~S.,  Kesteven M.~J.,  Haynes R.~F.,  Wellington K.~J.,
  Stewart R.~T.,  1986, \mn@doi [Monthly Notices of the Royal Astronomical
  Society] {10.1093/mnras/223.3.487}, 223, 487

\bibitem[\protect\citeauthoryear{Morlino, Blasi, Bandiera, Amato  \&
  Caprioli}{Morlino et~al.}{2013}]{Morlino2013}
Morlino G.,  Blasi P.,  Bandiera R.,  Amato E.,   Caprioli D.,  2013, \mn@doi
  [The Astrophysical Journal] {10.1088/0004-637X/768/2/148}, 768, 148

\bibitem[\protect\citeauthoryear{Newville, Stensitzki, Allen  \&
  Ingargiola}{Newville et~al.}{2014}]{Newville2014}
Newville M.,  Stensitzki T.,  Allen D.~B.,   Ingargiola A.,  2014, {LMFIT:
  Non-Linear Least-Square Minimization and Curve-Fitting for Python},
  \mn@doi{10.5281/zenodo.11813}

\bibitem[\protect\citeauthoryear{{Origlia} et~al.,}{{Origlia}
  et~al.}{2016}]{Origlia2016}
{Origlia} L.,  et~al., 2016, \mn@doi [\aap] {10.1051/0004-6361/201526649},
  \href {https://ui.adsabs.harvard.edu/abs/2016A&A...585A..14O} {585, A14}

\bibitem[\protect\citeauthoryear{Pandas~development team}{Pandas~development
  team}{2020}]{reback2020pandas}
Pandas~development team T.,  2020, {pandas-dev/pandas: Pandas},
  \mn@doi{10.5281/zenodo.3509134}, \url
  {https://doi.org/10.5281/zenodo.3509134}

\bibitem[\protect\citeauthoryear{{Parker} et~al.,}{{Parker}
  et~al.}{2005}]{Parker2005}
{Parker} Q.~A.,  et~al., 2005, \mn@doi [\mnras]
  {10.1111/j.1365-2966.2005.09350.x}, \href
  {https://ui.adsabs.harvard.edu/abs/2005MNRAS.362..689P} {362, 689}

\bibitem[\protect\citeauthoryear{Pastorello et~al.,}{Pastorello
  et~al.}{2004}]{Pastorello2004}
Pastorello A.,  et~al., 2004, \mn@doi [Monthly Notices of the Royal
  Astronomical Society] {10.1111/j.1365-2966.2004.07173.x}, 347, 74

\bibitem[\protect\citeauthoryear{Pavlovic, Dobardzic, Vukotic  \&
  Urosevic}{Pavlovic et~al.}{2014}]{Pavlovic2014}
Pavlovic M.,  Dobardzic A.,  Vukotic B.,   Urosevic D.,  2014, \mn@doi [Serbian
  Astronomical Journal] {10.2298/SAJ1489025P}, pp 25--40

\bibitem[\protect\citeauthoryear{Perek \& Kohoutek}{Perek \&
  Kohoutek}{1967}]{Perek1967}
Perek L.,  Kohoutek L.,  1967, {Catalogue of Galactic Planetary Nebulae}.
Publication House Czechoslovak Academy of Sciences

\bibitem[\protect\citeauthoryear{{Ritter}, {Parker}, {Lykou}, {Zijlstra},
  {Guerrero}  \& {Le D{\^u}}}{{Ritter} et~al.}{2021}]{Ritter2021}
{Ritter} A.,  {Parker} Q.~A.,  {Lykou} F.,  {Zijlstra} A.~A.,  {Guerrero}
  M.~A.,   {Le D{\^u}} P.,  2021, \mn@doi [\apjl] {10.3847/2041-8213/ac2253},
  \href {https://ui.adsabs.harvard.edu/abs/2021ApJ...918L..33R} {918, L33}

\bibitem[\protect\citeauthoryear{Robitaille}{Robitaille}{2019}]{Robitaille2019}
Robitaille T.,  2019, {APLpy v2.0: The Astronomical Plotting Library in
  Python}, \mn@doi{10.5281/zenodo.2567476}, \url
  {https://doi.org/10.5281/zenodo.2567476}

\bibitem[\protect\citeauthoryear{{Sabbadin}, {Minello}  \&
  {Bianchini}}{{Sabbadin} et~al.}{1977}]{Sab1977}
{Sabbadin} F.,  {Minello} S.,   {Bianchini} A.,  1977, \aap, \href
  {https://ui.adsabs.harvard.edu/abs/1977A&A....60..147S} {60, 147}

\bibitem[\protect\citeauthoryear{Samarakoon}{Samarakoon}{2018}]{Samarakoon2018}
Samarakoon B.,  2018, \mn@doi [KDU Journal of Multidisciplinary Studies (KJMS)]
  {10.13140/RG.2.2.17325.82404}, 1, 1

\bibitem[\protect\citeauthoryear{Sanduleak}{Sanduleak}{1976}]{Sanduleak1976}
Sanduleak N.,  1976, Publications of the Warner \& Swasey Observatory, 2, 57

\bibitem[\protect\citeauthoryear{{Schaefer}}{{Schaefer}}{2023}]{Schaefer2023}
{Schaefer} B.~E.,  2023, \mn@doi [\mnras] {10.1093/mnras/stad717}, \href
  {https://ui.adsabs.harvard.edu/abs/2023MNRAS.523.3885S} {523, 3885}

\bibitem[\protect\citeauthoryear{{Schirmer}}{{Schirmer}}{2013}]{Schirmer2013}
{Schirmer} M.,  2013, \mn@doi [\apjs] {10.1088/0067-0049/209/2/21}, \href
  {https://ui.adsabs.harvard.edu/abs/2013ApJS..209...21S} {209, 21}

\bibitem[\protect\citeauthoryear{{Smith}}{{Smith}}{2014}]{Smith2014}
{Smith} N.,  2014, \mn@doi [\araa] {10.1146/annurev-astro-081913-040025}, \href
  {https://ui.adsabs.harvard.edu/abs/2014ARA&A..52..487S} {52, 487}

\bibitem[\protect\citeauthoryear{{Sutherland} \& {Dopita}}{{Sutherland} \&
  {Dopita}}{1995}]{sut95}
{Sutherland} R.~S.,  {Dopita} M.~A.,  1995, \mn@doi [\apj] {10.1086/175180},
  \href {https://ui.adsabs.harvard.edu/abs/1995ApJ...439..365S} {439, 365}

\bibitem[\protect\citeauthoryear{{Tody}}{{Tody}}{1986}]{Tody1986}
{Tody} D.,  1986, in {Crawford} D.~L.,  ed.,  Society of Photo-Optical
  Instrumentation Engineers (SPIE) Conference Series Vol. 627, Instrumentation
  in astronomy VI. p.~733, \mn@doi{10.1117/12.968154}

\bibitem[\protect\citeauthoryear{Vink}{Vink}{2016}]{Vink2016}
Vink J.,  2016, \mn@doi [Handbook of Supernovae]
  {10.1007/978-3-319-21846-5_49}, pp 139--160

\bibitem[\protect\citeauthoryear{Vink}{Vink}{2020}]{Vink2020Book}
Vink J.,  2020, {Physics and Evolution of Supernova Remnants}.
Astronomy and Astrophysics Library, Springer International Publishing, Cham,
  \mn@doi{10.1007/978-3-030-55231-2}, \url
  {http://link.springer.com/10.1007/978-3-030-55231-2}

\bibitem[\protect\citeauthoryear{Vink \& Yamazaki}{Vink \&
  Yamazaki}{2013}]{Vink2013}
Vink J.,  Yamazaki R.,  2013, \mn@doi [The Astrophysical Journal]
  {10.1088/0004-637X/780/2/125}, 780, 125

\bibitem[\protect\citeauthoryear{Vink, Yamazaki, Helder  \& Schure}{Vink
  et~al.}{2010}]{Vink2010}
Vink J.,  Yamazaki R.,  Helder E.~A.,   Schure K.~M.,  2010, \mn@doi [The
  Astrophysical Journal] {10.1088/0004-637X/722/2/1727}, 722, 1727

\bibitem[\protect\citeauthoryear{Virtanen et~al.,}{Virtanen
  et~al.}{2020}]{Virtanen2020}
Virtanen P.,  et~al., 2020, \mn@doi [Nature Methods]
  {10.1038/s41592-019-0686-2}, 17, 261

\bibitem[\protect\citeauthoryear{Zhou, Vink, Li  \& Dom{\v{c}}ek}{Zhou
  et~al.}{2018}]{Zhou2018}
Zhou P.,  Vink J.,  Li G.,   Dom{\v{c}}ek V.,  2018, \mn@doi [The Astrophysical
  Journal] {10.3847/2041-8213/aae07d}, 865, L6

\bibitem[\protect\citeauthoryear{van~den Bergh}{van~den
  Bergh}{1978}]{VandenBergh1978}
van~den Bergh S.,  1978, \mn@doi [The Astrophysical Journal Supplement Series]
  {10.1086/190549}, 38, 119

\makeatother
\end{thebibliography}



\bsp	
\label{lastpage}
\end{document}